\documentclass[aps,prb,twocolumn,letterpaper]{revtex4-1}
\pdfoutput=1
\usepackage{amsmath,amssymb,amsfonts,upgreek}
\usepackage{color}
\usepackage{graphicx}
\usepackage{setspace}

\DeclareMathOperator{\Tr}{Tr} 

\begin{document}

\title{Strain-induced insulator-to-metal transition in LaTiO$_3$
  within DFT+DMFT}

\date{\today}

\author{Krzysztof Dymkowski}
\email{krzysztof.dymkowski@mat.ethz.ch} 
\affiliation{Materials Theory, ETH Z\"urich, Wolfgang-Pauli-Strasse
  27, 8093 Z\"urich, Switzerland} 
\author{Claude Ederer} 
\email{claude.ederer@mat.ethz.ch}
\affiliation{Materials Theory, ETH Z\"urich, Wolfgang-Pauli-Strasse
  27, 8093 Z\"urich, Switzerland} 

\begin{abstract}
We present results of combined density functional theory plus
dynamical mean-field theory (DFT+DMFT) calculations, which show that
the Mott insulator LaTiO$_3$ undergoes an insulator-to-metal
transition under compressive epitaxial strain of about $-2$\,\%. This
transition is driven by strain-induced changes in the crystal-field
splitting between the Ti $t_{2g}$ orbitals, which in turn are
intimately related to the collective tilts and rotations of the oxygen
octahedra in the orthorhombically distorted $Pbnm$ perovskite
structure. An accurate treatment of the underlying crystal structure
is therefore crucial for a correct description of the observed
metal-insulator transition. Our theoretical results are consistent
with recent experimental observations, and demonstrate that metallic
behavior in heterostructures of otherwise insulating materials can
emerge also from mechanisms other than genuine interface effects.
\end{abstract}
\pacs{}

\maketitle


Emergent phenomena at oxide interfaces are currently attracting a lot
of attention both from basic science as well as due to their prospects
for future technological
devices.\cite{Mannhart/Schlom:2010,Chakhalian/Millis/Rondinelli:2012,Hwang_et_al:2012}
Oxide heterostructures often exhibit properties that are not present
in the individual materials. For example, metallic conductivity has
been observed in heterostructures consisting of ultrathin layers of
LaTiO$_3$ (1-4 unit cells thick) embedded in
SrTiO$_3$,\cite{Ohtomo_et_al:2002} even though the corresponding bulk
materials are both insulators. This has been explained by charge
transfer between the Ti$^{3+}$ and Ti$^{4+}$ cations at the interfaces
between the two
components.\cite{Okamoto/Millis:2004,Okamoto/Millis/Spaldin:2006}
However, more recently, metallicity has also been found in thin films
of LaTiO$_3$ (15-45 nm thick) grown epitaxially on
SrTiO$_3$.\cite{Wong_et_al:2010, He_et_al:2012} It has been shown that
the sheet carrier density in these films scales linearly with film
thickness, which suggests that the metallicity is not just restricted
to the interface region. Furthermore, a comparison of LaTiO$_3$ films
grown on different substrates, suggests that the metallic properties
are controlled by substrate-induced epitaxial
strain.\cite{Wong_et_al:2010}

To better understand the physical mechanisms behind emerging
properties in oxide heterostructures, it is important to clearly
distinguish the effects of different factors, such as e.g. strain,
defects, structural and electronic reconstruction at the interface,
etc. Here, we present results of first principles calculations using a
combination of density functional theory and dynamical mean-field
theory
(DFT+DMFT),\cite{Georges_et_al:1996,Anisimov_et_al:1997,Kotliar/Vollhardt:2004,Held:2007}
assessing specifically the effect of epitaxial strain on LaTiO$_3$,
independent from interface effects or variations in defect
concentration. We find that compressive epitaxial strain can indeed
induce a metal-insulator transition, consistent with the experimental
observations, whereas tensile strain strongly enforces the insulating
character. Furthermore, we show that this behavior is controlled
mainly by strain-induced changes in the crystal-field splitting within
the Ti $t_{2g}$ bands.

\begin{figure}
\includegraphics[width=0.5\textwidth,clip]{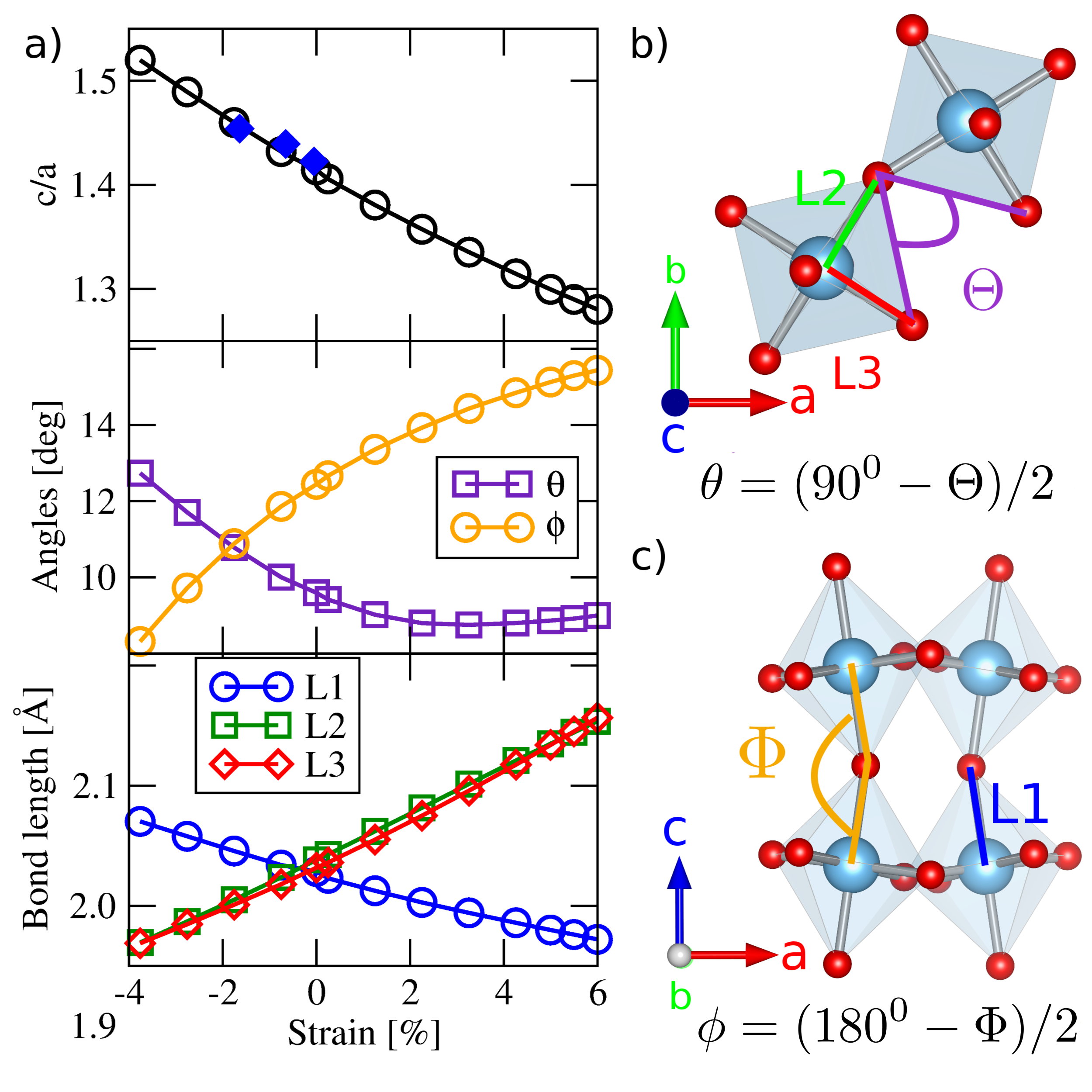}
\caption{a) Calculated $c/a$ ratio (upper panel), octahedral tilt
  angles (middle panel), and Ti-O bond distances (lower panel) in
  LaTiO$_3$ as function of in-plane strain. The $c/a$ ratio is
  calculated from the orthorhombic lattice parameters ($c/a \approx
  \sqrt{2}$ for zero strain), and is compared to experimental data
  from Ref.~\onlinecite{Wong_et_al:2010} (filled diamonds). b) and c)
  show projections of the orthorhombically distorted $Pbnm$ perovskite
  structure. The three different Ti-O bond distances are denoted as
  L1-L3.  The angles $\theta$ and $\phi$ measure in-plane
  ``rotations'' and out-of-plane ``tilts'', respectively, and are
  related to specific bond angles which are indicated in b) and c) by
  capital letters $\Theta$ and $\Phi$.}
\label{fig:struct}
\end{figure}

Bulk LaTiO$_3$ at room temperature is a paramagnetic Mott
insulator.\cite{Fujimori_et_al:1992,Arima/Tokura/Torrance:1993} The
theoretical description of such systems is challenging, since the
standard local density or generalized gradient approximations are not
suitable for describing the Mott-insulating state without any
symmetry-breaking long-range order.\cite{Georges_et_al:1996} However,
a good description of LaTiO$_3$ and other perovskite systems with
$d^1$ electron configuration of the transition metal cation can be
achieved within the DFT+DMFT
approach.\cite{Pavarini_et_al:2004,Craco_et_al:2004,Pavarini_et_al:2005}
Using this method, it has been shown that the Mott-insulating
character of $d^1$ perovskites with $Pbnm$ symmetry, such as
LaTiO$_3$,\cite{MacLean/Ng/Greedan:1979,Cwik_et_al:2003} is controlled
by the amplitude of the characteristic structural distortion,
i.e. tilts of the oxygen octahedra, the so-called GdFeO$_3$-type
distortion (see Fig.~\ref{fig:struct}b and
c).\cite{Pavarini_et_al:2004}

On the other hand, first principles calculations for a number of
$Pbnm$ perovskites have shown that epitaxial strain generally affects
both character and amplitude of the octahedral tilts (see
Ref.~\onlinecite{Rondinelli/Spaldin:2011} for a review). Since, as
stated above, the insulating character of LaTiO$_3$ is controlled by
the octahedral tilts, it can be expected that epitaxial strain will
have a pronounced effect on the electronic properties of LaTiO$_3$ and
eventually even drive the system towards a metallic state. Conversely,
in order to understand the effect of epitaxial strain on LaTiO$_3$ it
is thus very important to correctly account for the resulting changes
in the octahedral tilt distortion.


We therefore start by performing accurate structural relaxations for
LaTiO$_3$ under different epitaxial constraints in the following way.
We fix the two ``in-plane'' lattice constants to $a=b=(\epsilon+1)
\cdot a_0$, where the value $a_0=5.60$\,\AA\ corresponds to the
``coherent structure'',\cite{Rondinelli/Spaldin:2011} i.e. minimal
energy under the constraint $a=b$. For each strain $\epsilon$ we then
relax the ``out-of-plane'' lattice constant $c$ and all internal
structural parameters (for simplicity we assume that the films grow
with the orthorhombic $c$ direction (long axis) perpendicular to the
substrate plane). All structural relaxations are performed within the
generalized gradient approximation (GGA)
\cite{Perdew/Burke/Ernzerhof:1996} to density functional theory (DFT)
using the Quantum ESPRESSO code \cite{Giannozzi_et_al:2009} with a
plane-wave basis and ultrasoft pseudopotentials.\cite{Vanderbilt:1990}
Since our work focuses on the room-temperature paramagnetic phase of
LaTiO$_3$, all calculations are performed for the non-spin-polarized
case. Further technical details can be found in the supplemental
material.

We note that the octahedral tilt distortion in perovskites is assumed
to be driven by the size ratio between the two different cations
\cite{Goldschmidt:1926} as well as by hybridization between the oxygen
anions and the large $A$-site cation.\cite{Pavarini_et_al:2004} These
effects are generally well described within GGA. Thus, an accurate
treatment of electronic correlations is not necessary to describe the
structural properties of LaTiO$_3$. Nevertheless, in order to
benchmark the accuracy of our calculated structural parameters, we
first perform a full unconstrained structural relaxation of the
orthorhombic $Pbnm$ bulk structure. All calculated lattice parameters
agree well with available experimental data. In particular, the
octahedral tilt angles deviate by less than $1^\circ$ from the
experimental values (see supplemental information for more details).


We now discuss the results of our structural relaxations under
epitaxial strain. Fig.~\ref{fig:struct}a shows the calculated $c/a$
ratio, the octahedral tilt angles, and the Ti-O bond distances as a
function of applied strain. As expected, compressive (tensile) strain
results in an elongation (reduction) of $c$. The calculated $c/a$
shows good agreement with experimental data from
Ref.~\onlinecite{Wong_et_al:2010}.  Furthermore, it can be seen that
epitaxial strain has a pronounced effect on the octahedral tilt angles
as well as on the Ti-O bond distances. The octahedral tilt angles show
the expected behavior, similar to what has been observed in other
$Pbnm$ perovskites.\cite{Rondinelli/Spaldin:2011} The in-plane
``rotations'', characterized by the angle $\theta$, increase under
compressive strain, in order to decrease the unit cell area within the
$a$-$b$ plane. On the other hand, the out-of-plane ``tilts'',
characterized by $\phi$, decrease due to the elongation along $c$,
which straightens out the Ti-O-Ti bonds along this direction. Tensile
strain leads to the opposite trends. When looking at the Ti-O bond
distances, it can be seen that all three bond distances are nearly
equal for the unstrained structure, i.e. there is no significant
Jahn-Teller distortion of the oxygen octahedra. Under strain the
lengths of the two ``in-plane'' bonds (L2/L3) remain approximately
equal and are reduced (elongated) for compressive (tensile)
strain. The bond distance along $c$ (L1) exhibits the opposite trend.

Thus, while the octahedral tilt angles follow the trends expected for
a network of rigid octahedra under strain, the octahedra are in fact
not completely rigid and also deform as a result of the applied
strain. Part of the strain is therefore compensated by changes in the
octahedral tilt angles and part of it is compensated by the
deformation of the octahedra. The deviation from the behavior of rigid
octahedra is especially pronounced for large tensile strain ($\sim$
4-6\,\%) where the angle $\theta$ becomes nearly constant and even
shows a slight increase with increasing strain (see
Fig.~\ref{fig:struct}a).


After analyzing the structural changes under strain, we now turn to
the electronic properties. Following previous
work,\cite{Pavarini_et_al:2004,Craco_et_al:2004,Pavarini_et_al:2005}
we employ a DMFT treatment to account for the electron-electron
interaction within the partially filled bands around the Fermi energy
with predominant Ti $t_{2g}$ character. We express the Kohn-Sham
Hamiltonian in a basis of maximally localized Wannier functions
\cite{Marzari_et_al:2012,Mostofi_et_al:2008} describing the effective
Ti $t_{2g}$ bands, and use this as the noninteracting part $H_0$ of a
multiband Hubbard Hamiltonian $H=H_0+H_\text{int}$. The local
electron-electron interaction $H_\text{int}$ is expressed in the
Slater-Kanamori form, including both spin-flip and pair hopping terms,
with the parameters $U$ and $J$ describing the inter-orbital Coulomb
interaction and Hund's rule coupling, respectively (see
e.g. Ref.~\onlinecite{Kovacik_et_al:2012}). The local Green's function
is then calculated within DMFT \cite{Georges_et_al:1996} at
temperature $T=1/(k_\text{B}\beta)$ using a continuous time
hybridization expansion quantum Monte-Carlo solver
\cite{Gull_et_al:2011} implemented in the TRIQS 0.9 code.\cite{triqs}
Off-diagonal elements between different orbitals on the same site are
taken into account both for the local Green's functions and the
self-energy. Such off-diagonal elements appear due to the
symmetry-lowering associated with the octahedral tilts. No sign
problem related to these off-diagonal elements was encountered in the
Monte-Carlo calculations. All calculations are performed for
$\beta=40$\,eV$^{-1}$ ($T=290$\,K) and $J=0.64$\,eV, while $U$ is
varied between 4\,eV and 5.9\,eV. More details can be found in the
supplemental material.


\begin{figure}
\includegraphics[width=0.6\columnwidth]{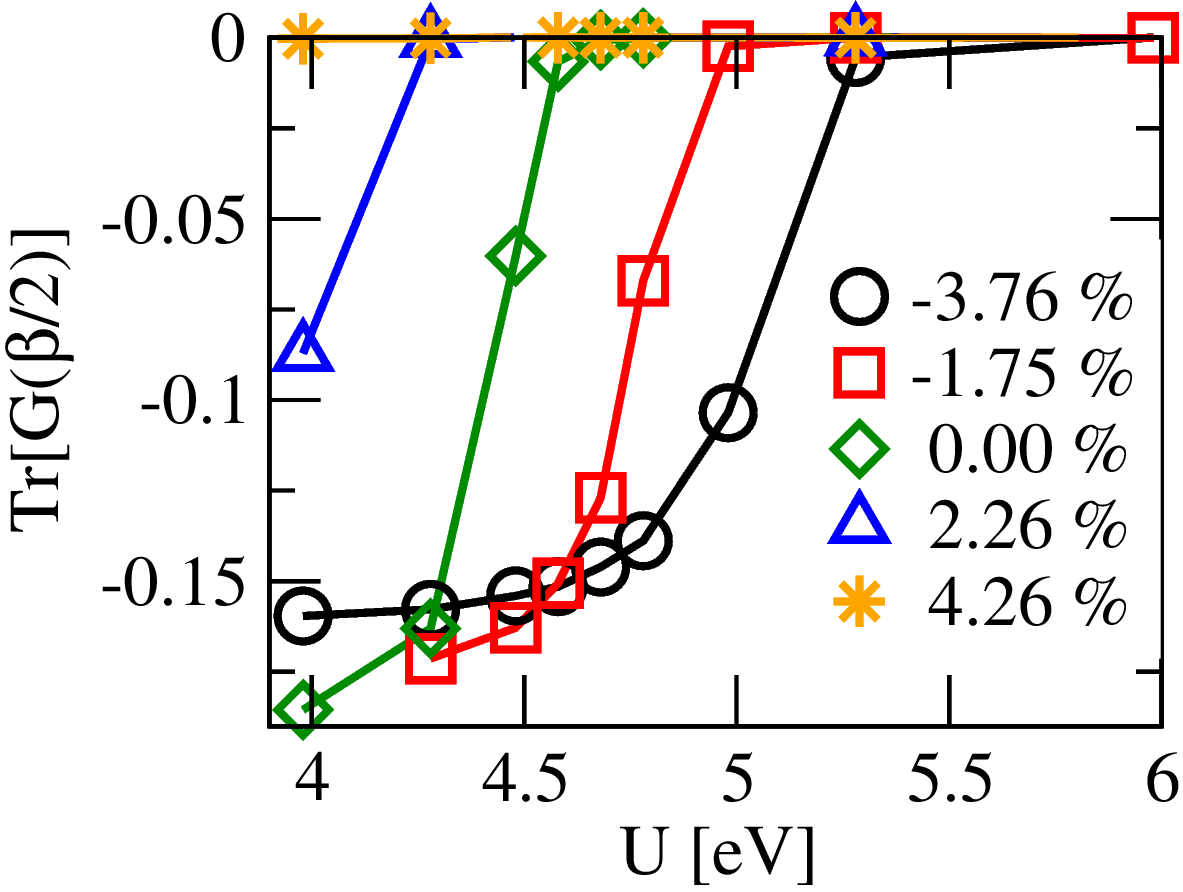}
\caption{Trace of the the local Green's function at $\tau=\beta/2$ as
  function of the interaction parameter $U$, calculated for different
  values of epitaxial strain.}
\label{fig:glocal}
\end{figure}

Fig.~\ref{fig:glocal} shows the trace of the local Green's function at
imaginary time $\tau = \beta/2$ as a function of $U$,\footnote{Note
  that we follow the usual convention of presenting all results as
  function of the intra-orbital Hubbard interaction $U$, even though
  the physically relevant quantity determining the metal-insulator
  transition in a three-orbital model with one or two electrons is
  $U'-J=U-3J$, as shown e.g. in
  Ref.~\onlinecite{Werner/Gull/Millis:2009}.} calculated for different
strain values. Based on the relation \mbox{$A(\omega = 0) =
  -\tfrac{1}{\pi} \lim_{\beta \rightarrow \infty} \beta G(\beta/2)$},
this can be taken as a measure of the total spectral function at the
Fermi level (see e.g. Ref.~\onlinecite{Kovacik_et_al:2012}). It can be
seen, that there is a change from \mbox{$\Tr G(\beta/2) \neq 0$} to
\mbox{$\Tr G(\beta/2) \approx 0$} in the interval \mbox{$4\,\text{eV}
  < U < 5.5\,\text{eV}$} for all negative strain values, i.e. the
system undergoes a metal-insulator transition with increasing $U$. It
can further be seen, that epitaxial strain has a pronounced effect on
the critical $U$ for this transition. While tensile strain strongly
decreases the critical $U$ (to values smaller than 4\,eV for $\epsilon
> 2.26$\,\%), compressive strain has the opposite effect, and
increases the critical $U$ to slightly above 5\,eV at $-3.76$\,\%
strain. Thus, compressive (tensile) strain favors the metallic
(insulating) state.

Thus, irrespective of any uncertainty regarding the appropriate value
of $U$ and possible small inaccuracies in the calculated GGA lattice
parameters, our calculations clearly show that compressive strain
promotes metallicity in LaTiO$_3$, consistent with the experimental
observation of metallic conductivity in thin films of compressively
strained LaTiO$_3$.\cite{Wong_et_al:2010} Assuming a realistic $U$ for
the Ti $t_{2g}$ Wannier orbitals in the range of 4.5-5\,eV, our
results show that a strain-induced insulator-to-metal transition
occurs for compressive strains around $-2$\,\%, which compares well
with the lattice mismatch of $-1.6$\,\% for LaTiO$_3$ films grown on
SrTiO$_3$. Of course our calculations do not exclude the possibility
that other aspects such as oxygen vacancies or interface polarity can
also strongly influence the conductivity of the LaTiO$_3$ films.


\begin{figure}
\begin{center}
\raisebox{1.6cm}{a)}
\begin{minipage}[c]{0.8\columnwidth}
\includegraphics[height=0.55\textwidth]{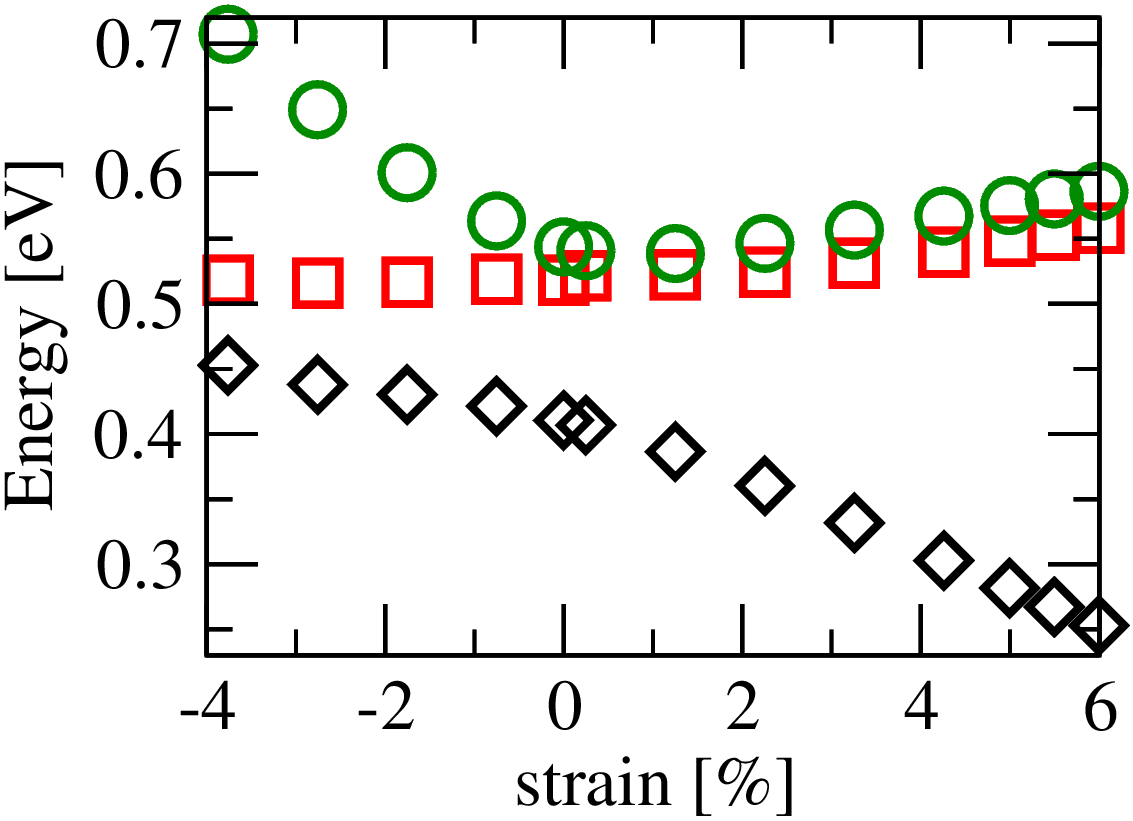}
\end{minipage}

\smallskip

\raisebox{1.6cm}{b)}
\begin{minipage}[c]{0.8\columnwidth}
\includegraphics[height=0.55\textwidth]{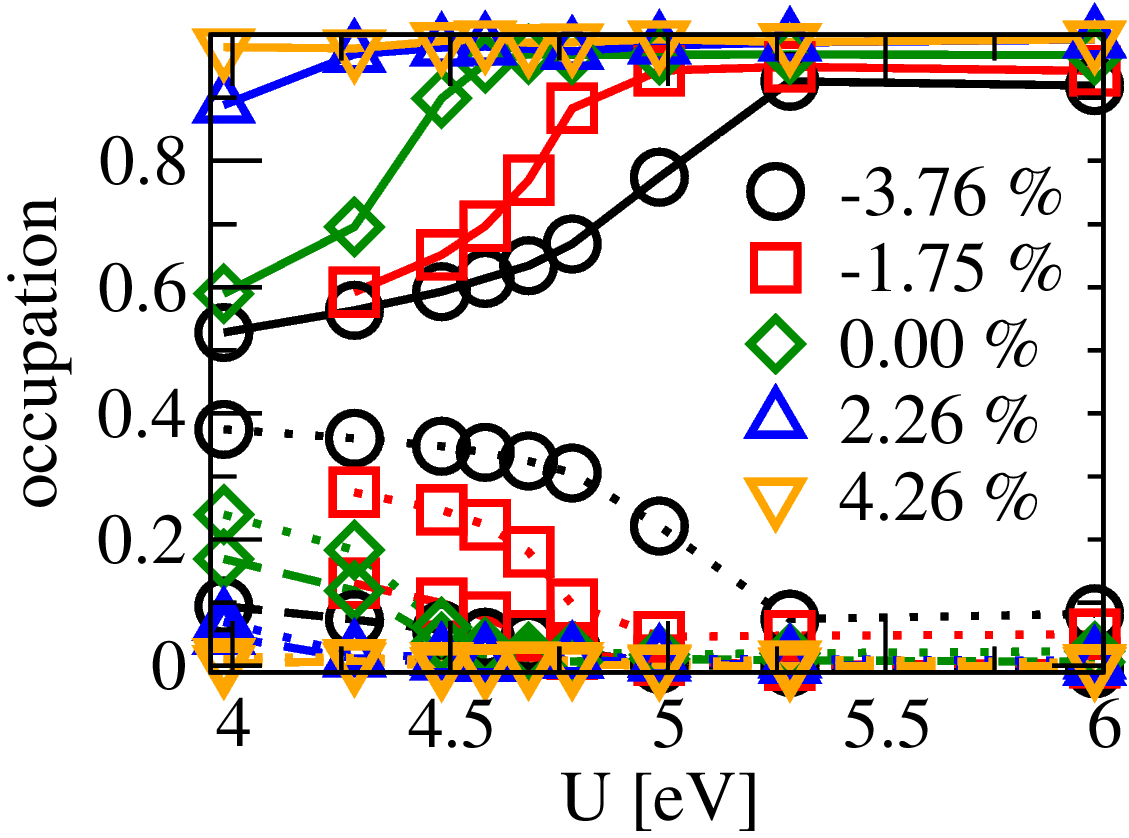}
\end{minipage}

\smallskip

\raisebox{1.6cm}{c)}
\begin{minipage}[c]{0.8\columnwidth}
\hspace*{5mm}\includegraphics[height=0.55\textwidth]{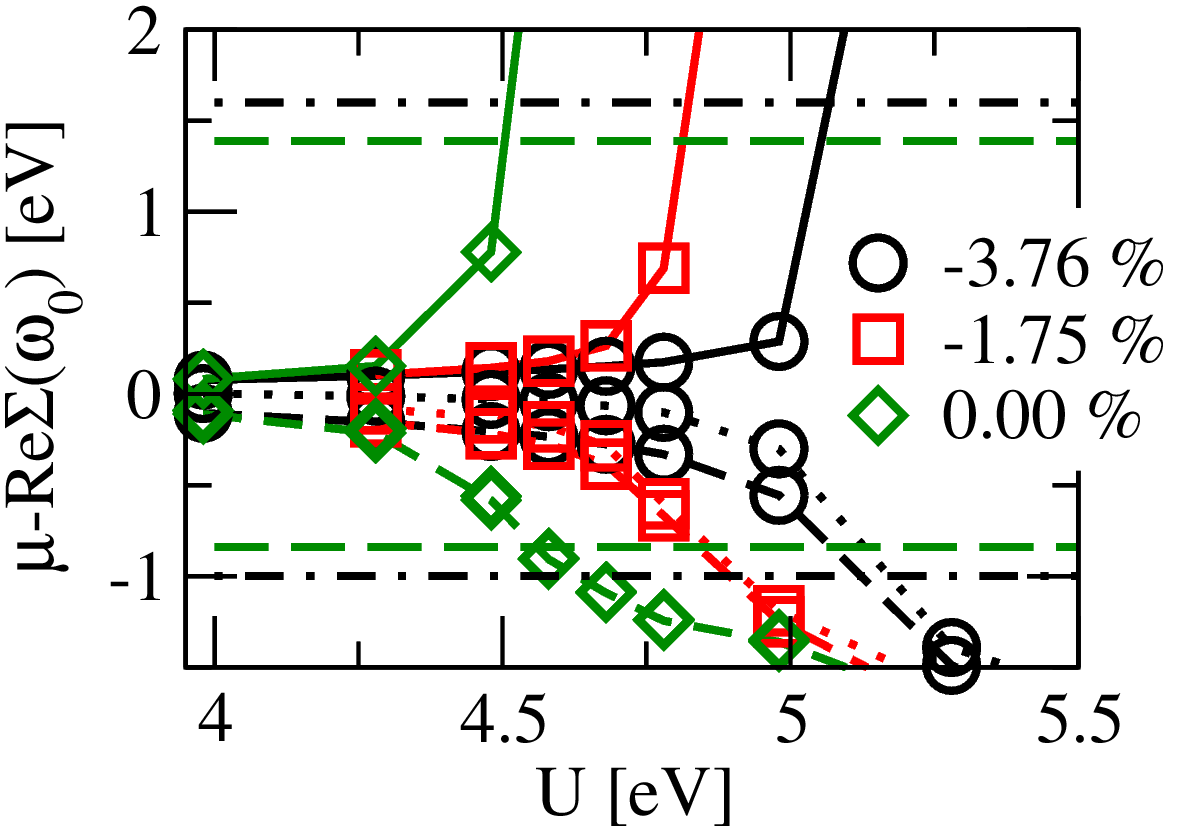}
\end{minipage}
\end{center}
\caption{a) Strain dependence of the three ``crystal-field levels''
  corresponding to the effective Ti $t_{2g}$ Wannier orbitals. b) and
  c) Occupations and ``effective'' chemical potential $\mu - {\rm Re}
  \Sigma(\omega_0)$, respectively, as function of the interaction
  parameter $U$, calculated for different strain values. The three
  different orbitals for each strain are indicated by solid, dotted,
  and dashed lines, in order of decreasing occupation. The dashed and
  dot-dashed horizontal lines in c) indicate the band-edges of the
  non-interacting system for 0\,\% and 3.76\,\% strain,
  respectively. In a) and c) the Fermi level was used as zero energy
  reference for each strain value.}
\label{fig:orbital}
\end{figure}


We now analyze the strain-induced transition in more detail by looking
at the occupations and energy levels of the three individual $t_{2g}$
orbitals. Fig.~\ref{fig:orbital} shows the eigenvalues of the
occupation matrix $n_{mm'} = - \sum_\sigma G^\sigma_{mm'}(\beta)$
together with the ``crystal-field levels'' of the effective Ti
$t_{2g}$ Wannier functions. The latter are obtained by diagonalizing
the local on-site Hamiltonian $H_0$. We have verified that the
resulting ``crystal field basis'' is very similar to the basis
diagonalizing the local occupation matrix (the scalar product between
the eigenvector corresponding to the lowest crystal-field orbital and
the eigenvector corresponding to the ``nearly-filled'' orbital is
larger than 0.96 for all strain values).

It can be seen from Fig.~\ref{fig:orbital}a that in the unstrained
state one orbital is lower in energy by 108\,meV compared to the other
two $t_{2g}$ states, whereas the splitting between the two
energetically higher orbitals is rather small (26 meV). This is
consistent with the orbital splittings calculated in
Ref.~\onlinecite{Pavarini_et_al:2005} for bulk LaTiO$_3$. We point out
that this crystal-field splitting is a result of the octahedral
rotations, which lower the symmetry of the Ti sites to triclinic,
whereas the Ti-O bond distances are essentially equal along all three
directions (see Fig.~\ref{fig:struct}a). Tensile strain conserves the
energetic order between the $t_{2g}$ orbitals, i.e. one orbital at
lower energy and two nearly degenerate levels at higher energies, and
strongly increases the corresponding splitting. In contrast,
compressive strain removes the near degeneracy between the two
orbitals with higher energy, while reducing the splitting between the
lowest and the intermediate energy level to 65\,meV at a compressive
strain of $-3.76$\,\%.  Thus, applying strain does not simply increase
the overall crystal-field splitting, but leads to qualitative changes
in the relative energy differences between the three Ti $t_{2g}$
Wannier orbitals.

The changes in the crystal-field levels are mirrored in the orbital
occupations calculated within DMFT (see Fig.~\ref{fig:orbital}b). For
large tensile strain the energetically lowest orbital is essentially
completely filled, whereas the other two orbitals remain empty. With
increasing compressive strain one of the two empty orbitals gains some
occupation at the expense of the filled orbital. This trend can
clearly be seen from the occupations for large $U$, i.e. where the
system is insulating. The transition to the metallic phase is then
accompanied by a strong charge transfer from the ``nearly filled''
into the ``nearly empty'' orbital, while the third orbital remains
mostly unaffected.

It thus appears that the decrease of the splitting between the two
lowest crystal-field levels under compressive strain is crucial to
facilitate the metallic state in LaTiO$_3$. In contrast, tensile
strain strongly increases this crystal-field splitting and thus
enforces the insulating state, i.e. the critical $U$ for the
transition is strongly reduced (see Fig.~\ref{fig:glocal}). This is
further evidenced by the orbitally-resolved real part of the self
energy at zero frequency, which can be viewed as additional
``effective'' chemical potential shift (see
e.g. Ref.~\onlinecite{Keller_et_al:2004}). As can be seen from
Fig.~\ref{fig:orbital}c, where $\Sigma(0)$ has been approximated by its
value at the lowest Matsubara frequency, the electron-electron
interaction strongly enhances the splitting between the energetically
lowest and the two higher lying orbitals in the insultating regime and
effectively shifts the chemical potential beyond the boundaries of the
non-interacting bands.\cite{Keller_et_al:2004} It can also be seen
from Fig.~\ref{fig:orbital}c that on the other hand the $t_{2g}$
bandwidth is only weakly affected by epitaxial strain. This is a
consequence of two competing effects on the dominant nearest neighbor
hopping amplitudes: the shorter Ti-O bond distances within the $a$-$b$
plane under compressive strain would in principle lead to an increase
in hopping, on the other hand the greater distortion of the
corresponding Ti-O-Ti bond angle has exactly the opposite effect (see
supplemental information for more details). This suggests that the
tendency towards the metallic phase under compressive strain is mainly
controlled by strain-induced changes in the crystal-field energies,
which are strongly enhanced due to the electron-electron interaction.

\begin{figure}
\includegraphics[width=0.7\columnwidth]{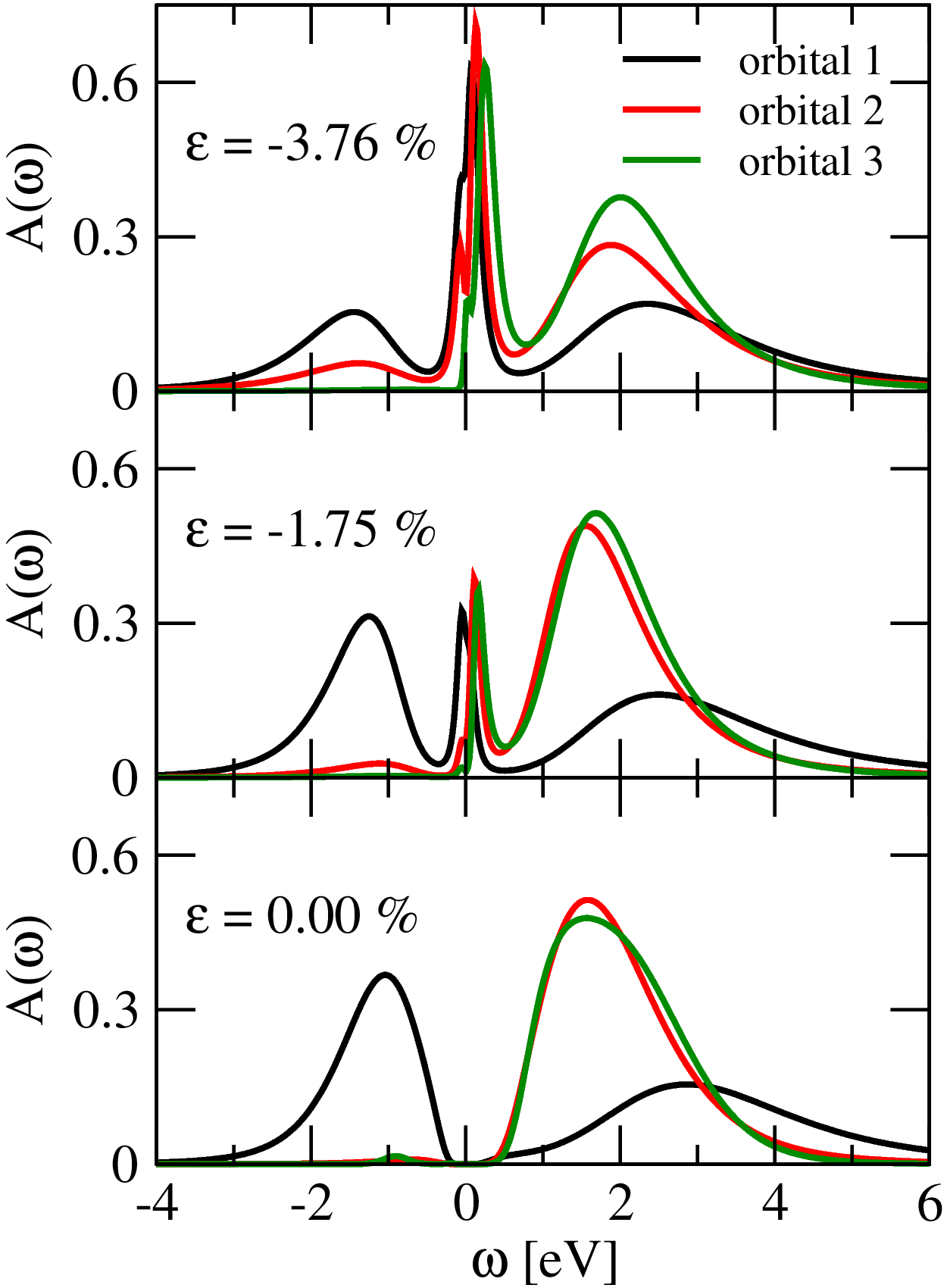}
\caption{Orbitally-resolved spectral functions for $U=4.78$\,eV and
  $\epsilon=0$ (lower panel), $\epsilon=-1.75$\,\% (middle panel), and
  $\epsilon=-3.76$\,\% (upper panel). Orbitals 1, 2, and 3 correspond
  to the eigenfunctions of the occupation matrix and are ordered with
  decreasing occupation.}
\label{fig:spectral}
\end{figure}

Finally, in Fig.~\ref{fig:spectral} we present orbitally-resolved
spectral functions for different strain values and $U=4.78$\,eV. The
spectral functions are obtained from the imaginary time Green's
functions by analytic continuation using the maximum entropy method
\cite{Jarrell/Gubernatis:1996} (see also supplemental material). The
spectral functions obtained in the insulating state for zero strain
agree well with those reported in
Ref.~\onlinecite{Pavarini_et_al:2005} for bulk LaTiO$_3$. All spectral
weight below zero energy corresponds to the occupied orbital with
lowest energy. With increasing compressive strain the systems becomes
metallic and strong quasiparticle features appear around zero energy
for all three orbitals, consistent with $\Tr G(\beta/2) \neq 0$ shown
in Fig.~\ref{fig:glocal}. Nevertheless, orbital 3 has negligible
spectral weight at negative energies for all strain values.

In summary, our calculations demonstrate an insulator-to-metal
transition under compressive epitaxial strain in LaTiO$_3$, consistent
with recent reports of metallic conductivity in LaTiO$_3$ films grown
on SrTiO$_3$. The origin of this transition is the reduced
crystal-field splitting under compressive strain, which facilitates
electron transfer between the two lowest Ti $t_{2g}$ levels. In
contrast, the increase of this splitting under tensile strain leads to
a strong reduction of the critical $U$ for the insulating state, which
suggests that the insulating character of LaTiO$_3$ could be enhanced
by growing it on substrates with slightly larger lattice constants.
We note that, while additional interface effects or the presence of
defects might further influence the properties of the real system, our
results indicate that such effects are not essential for explaining
the observed metallic behavior.

The fact that strain can alter the delicate balance between electron
hopping and Coulomb interaction, and destroy the Mott insulating state
in LaTiO$_3$, has been previously suggested.\cite{Ishida/Liebsch:2008}
However, the corresponding calculations were performed for a
simplified tetragonal crystal structure without octahedral tilts. Our
results demonstrate the close interplay between structure and
electronic properties in LaTiO$_3$ and also show that it is crucial to
consider realistic crystal structures within DMFT calculations.

It is clear that epitaxial strain is an important factor for the
emerging properties observed in oxide heterostructures. Therefore, in
order to understand the effect of the interface in these
heterostructures, the properties of the corresponding strained
material should be taken as reference, not the properties of the bulk
material. Interface effects can then successively be introduced using
slab supercells. For example a fully self-consistent DFT+DMFT study of
short period LaTiO$_3$-SrTiO$_3$ multilayer-structures, including full
structural relaxation of the unit cell, has been presented
recently.\cite{Lechermann/Boehnke/Grieger:2013} While no detailed
analysis of the origin of the metallicity (e.g. strain versus
interface effects) or of the degree of octahedral tilts has been
presented, Ref.~\onlinecite{Lechermann/Boehnke/Grieger:2013}
demonstrates the feasibility of such calculations even for complex
multilayer structures with large unit cells.

\begin{acknowledgments}
  This work was supported by ETH Z\"urich, the Swiss HP2C platform,
  and the Swiss National Science Foundation through the MaNEP centre
  and project grant no. 200021\_143265. We thank Michel Ferrero,
  Leonid Pourovskii, and Silke Biermann for useful discussions and
  their help with the TRIQS code.
\end{acknowledgments}

\bibliography{references}

\end{document}


\title{Supplemental Material --- Strain-induced insulator-to-metal
  transition in LaTiO$_3$ within DFT+DMFT}

\date{\today}

\author{Krzysztof Dymkowski} 
\email{krzysztof.dymkowski@mat.ethz.ch} 
\affiliation{Materials Theory, ETH Z\"urich, Wolfgang-Pauli-Strasse
  27, 8093 Z\"urich, Switzerland} 
\author{Claude Ederer} 
\email{claude.ederer@mat.ethz.ch}
\affiliation{Materials Theory, ETH Z\"urich, Wolfgang-Pauli-Strasse
  27, 8093 Z\"urich, Switzerland} 

\begin{abstract}
Here, we provide supplemental material for our paper ``Strain-induced
insulator-to-metal transition in LaTiO$_3$ within DFT+DMFT. We present
results of our structural relaxation for bulk (unconstrained) $Pbnm$
LaTiO$_3$ as well as the calculated Kohn-Sham bandstructures for
LaTiO$_3$ under different strain conditions. We highlight in
particular the projection of these bands on atomic Ti $t_{2g}$
orbitals. We also provide further details on the construction of our
maximally localized Wannier functions with predominant Ti $t_{2g}$
character, we present real-space pictures of the resulting Wannier
functions, and we give a brief discussion of the strain dependence of
the nearest neighbor hopping amplitudes. Finally, we provide a more
detailed description of the DMFT approach employed in our work.
\end{abstract}
\pacs{}

\maketitle

\paragraph*{Technical details of DFT calculations.}

As already stated in our article, all structural relaxations are
performed within the generalized gradient approximation (GGA)
\cite{Perdew/Burke/Ernzerhof:1996} to density functional theory (DFT)
using the Quantum ESPRESSO code \cite{Giannozzi_et_al:2009} with a
plane-wave basis and ultrasoft pseudopotentials.\cite{Vanderbilt:1990}
The La(5$s$,5$p$) and Ti(3$s$,3$p$) semicore states are included in
the valence. We use a $6 \times 6 \times 4$ k-point mesh and a plane
wave kinetic energy cutoff of $40$\,Ry for the wave-functions and
$480$\,Ry for the charge density. Internal structural degrees of
freedom are relaxed until all forces are smaller than
$10^{-4}$\,Ry/Bohr and the total energy is converged within
$10^{-5}$\,Ry. The out-of-plane lattice constant for the strained
structures is determined by performing a series of calculations where
all internal coordinates are relaxed using fixed lattice vectors, and
then varying the value of $c$ to find the minimal energy for fixed
in-plane constraint $a=b$ .

\paragraph*{Structural relaxation of bulk LaTiO$_3$.}

Table~\ref{table:bulk} lists the structural parameters we obtain by
performing a fully unconstrained structural relaxation for the $Pbnm$
bulk structure. All three lattice constants agree well with the
experimental data. Since we are performing nonmagnetic calculations,
it is most instructive to compare the calculated values to the
structural parameters measured at room temperature, where the system
is not magnetically ordered. The only noticeable difference between
calculated and measured values occurs for the lattice parameter $a$,
where the calculated value is about 1.8\,\% smaller than the measured
values. This could indicate a subtle effect of electronic correlations
on this lattice parameter, which shows a slightly unconventional
behavior, i.e. $a>b$ in the experimental structure, whereas a tilting
of rigid octahedra would lead to $a<b$ for the $a^-a^-c^+$ tilt
pattern realized in $Pbnm$ distorted
perovskites.\cite{Woodward:1997,Zayak_et_al:2006} This anomaly is
probably also related to the octahedral distortion (different edge
lengths) discussed in Ref.~\onlinecite{Cwik_et_al:2003} for
LaTiO$_3$. Nevertheless, the difference between calculated and
measured $a$ is still within the typical limits of DFT-based
calculations and provides sufficient accuracy for the present study.

\begin{table}
\caption{Orthorhombic lattice constants $a$, $b$, and $c$ of the
  $Pbnm$ structure of LaTiO$_3$ and octahedral tilt angles $\theta$
  and $\phi$ (defined in Fig.~1b and c of the main article). Our
  calculated values are compared with data from different experiments
  performed at different temperatures. }
\label{table:bulk}
\begin{ruledtabular}
\begin{tabular}{ccccc}
 & this work & Ref.~\onlinecite{Cwik_et_al:2003} & Ref.~\onlinecite{Cwik_et_al:2003} & Ref.~\onlinecite{Eitel/Greedan:1986} \\
 &  & $T=8$\,K & $T=293$\,K & $T=298$\,K \\
\hline
$a$ [\AA] & 5.53 & 5.64 & 5.63 & 5.62 \\
$b$ [\AA] & 5.62 & 5.59 & 5.62 & 5.61 \\
$c$ [\AA] & 7.87 & 7.90 & 7.91 & 7.92 \\
$\theta$ [$^\circ$] & 8.2 & 9.4 & 9.3 & 8.1 \\
$\phi$ [$^\circ$] & 12.2 & 13.1 & 12.9 & 11.3 
\end{tabular}
\end{ruledtabular}
\end{table}

The octahedral tilt and rotation angles show excellent agreement with
the experimental data. We note that there is also some disagreement
between the two different room temperature measurements.

\paragraph*{Electronic structure of strained LaTiO$_3$.}

\begin{figure}[hbtp!]
\includegraphics[width=0.65\columnwidth]{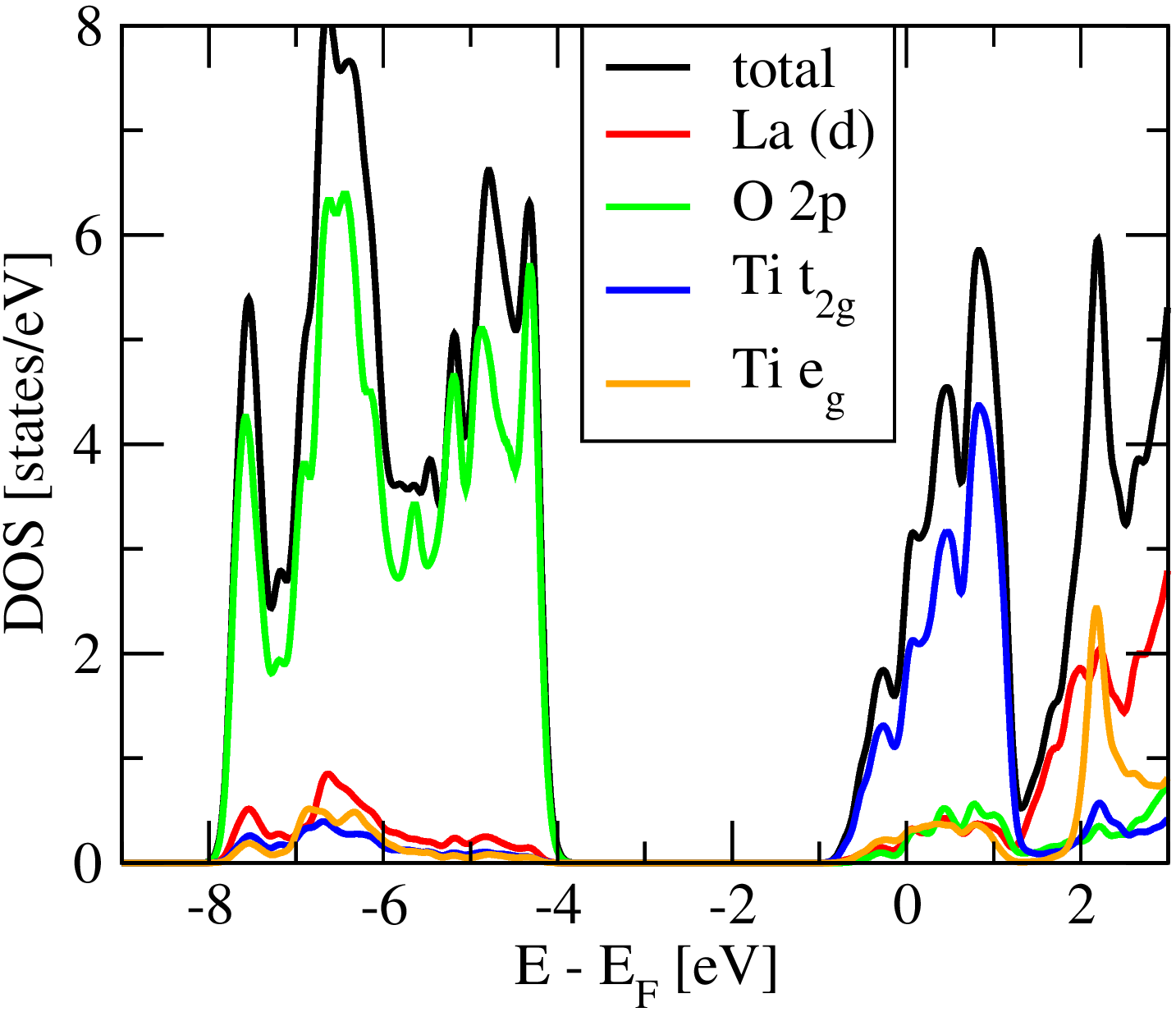}
\caption{Total and projected densities of states (DOS) per formula
  unit for nominally unstrained LaTiO$_3$ ($a=b=a_0$). Energies are
  expressed relative to the Fermi level $\epsilon_\text{F}$.}
\label{fig:dos}
\end{figure}

\begin{figure}
\raisebox{2cm}{a)}
\begin{minipage}[c]{0.8\columnwidth}
\includegraphics[width=0.9\textwidth]{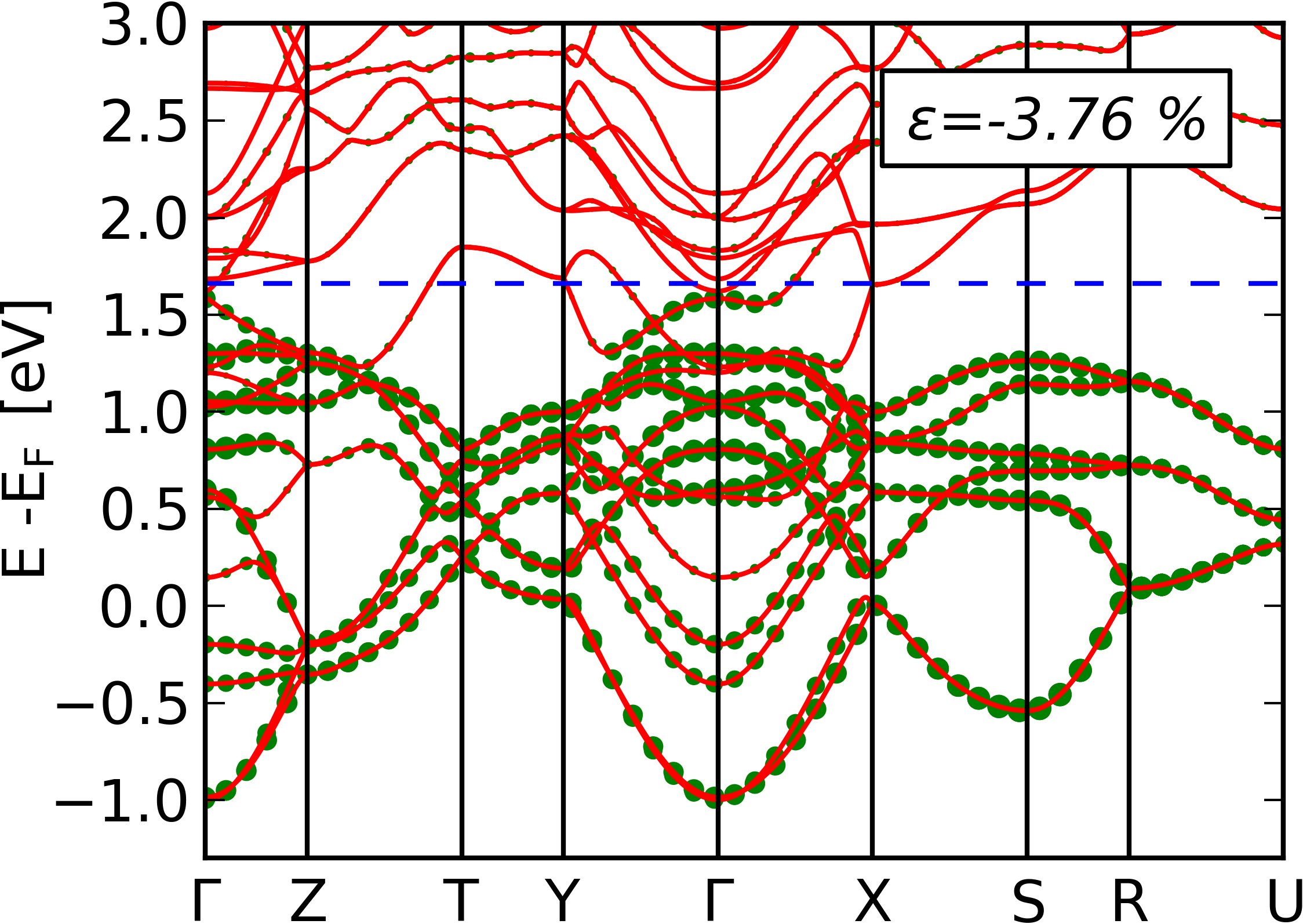}
\end{minipage}

\smallskip

\raisebox{2cm}{b)} 
\begin{minipage}[c]{0.8\columnwidth}
\includegraphics[width=0.9\textwidth]{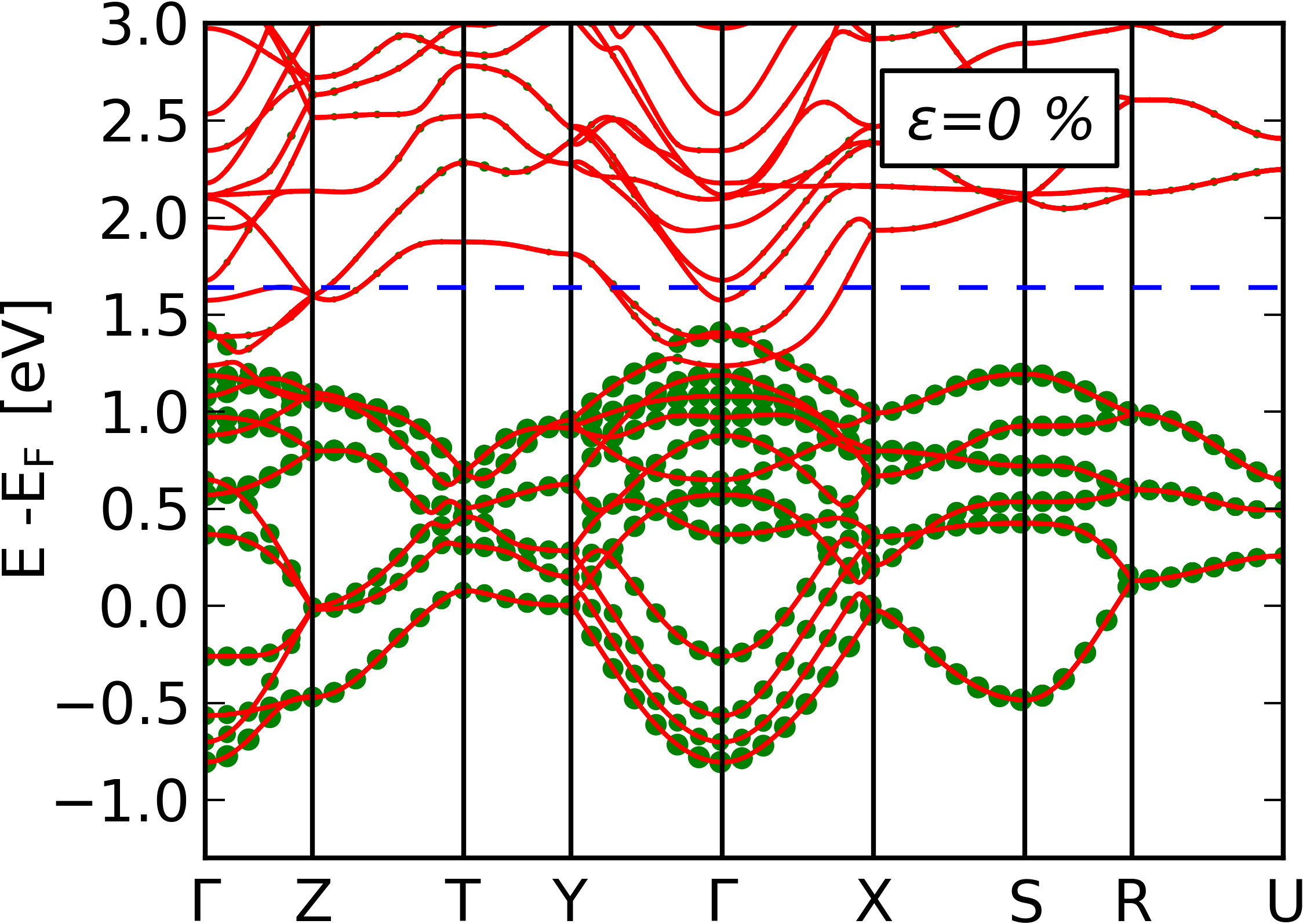}
\end{minipage}

\smallskip

\raisebox{2cm}{c)}
\begin{minipage}[c]{0.8\columnwidth}
\includegraphics[width=0.9\textwidth]{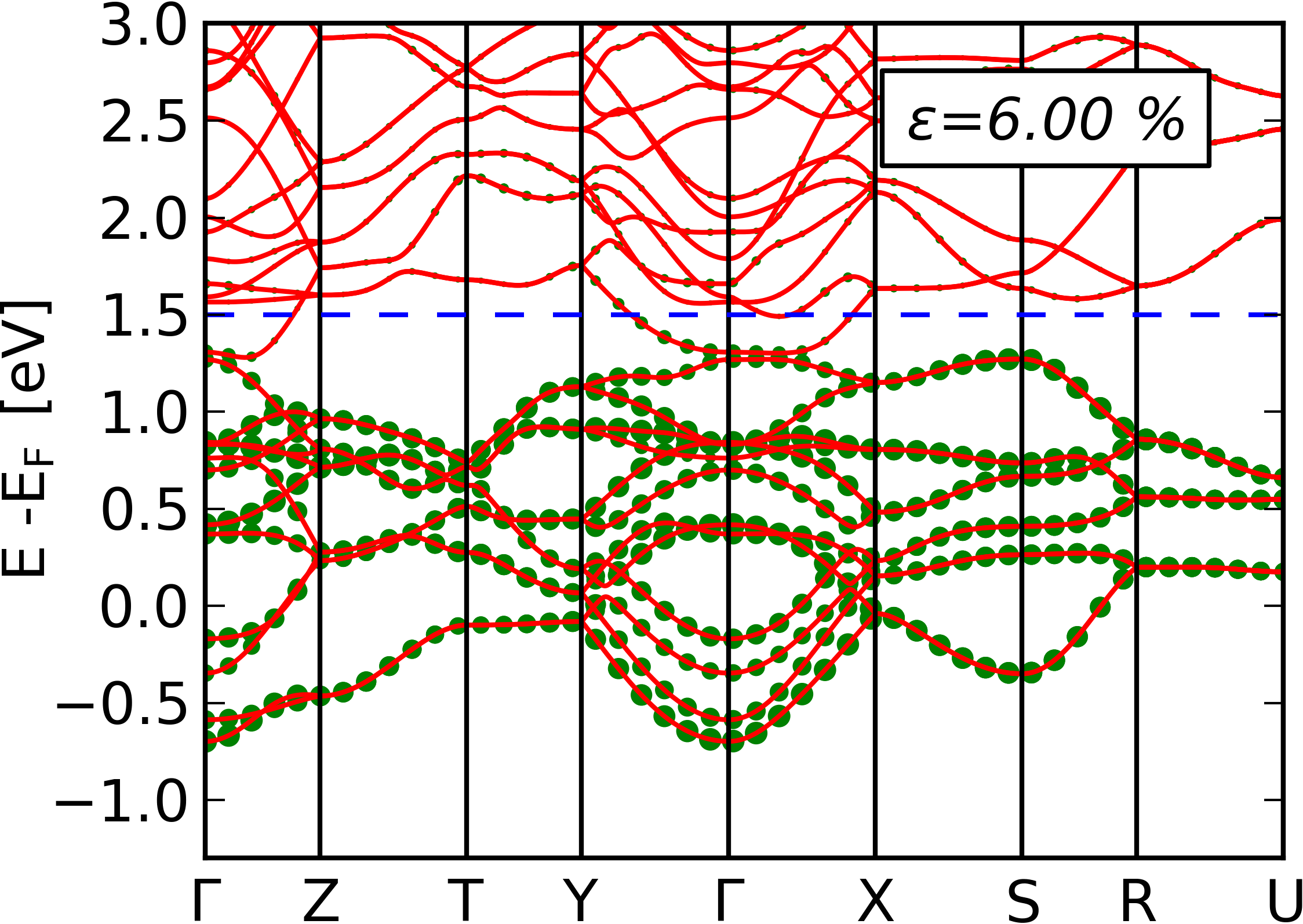}
\end{minipage}
\caption{Bandstructure around the Fermi level for $-$3.76\,\% strain
  (a), for the unstrained case (b), and for +6.0\,\% strain (c). The
  size of the green dots represents the Ti $t_{2g}$ character in the
  corresponding Bloch function. The dashed blue lines in a) and b)
  mark the upper boundaries of the energy window used to obtain the
  maximally localized Wannier functions (in c) there is a small gap
  between the bands with predominant Ti $t_{2g}$ character and other
  bands at higher energies). Energies are expressed relative to the
  Fermi level E$_\text{F}$.}
\label{fig:bands}
\end{figure}

Fig.~\ref{fig:dos} shows the total and projected densities of states
for nominally unstrained LaTiO$_3$, i.e. with $a=b=a_0$. The
constraint $a=b$ does not lead to any significant changes compared to
the fully unconstrained bulk case. It can be seen that the bands
around the Fermi level have predominant Ti $t_{2g}$ character and are
separated from bands at lower energy with predominant O $p$ character,
while they overlap slightly with La $d$ and Ti $e_{g}$ bands at higher
energies.

The Ti $t_{2g}$ character in the band-structure can also clearly be
identified from the ``fatbands'' shown in Fig.~\ref{fig:bands}. Here,
the Ti $t_{2g}$ character of the various bands is visualized by the
green dots, the size of which is proportional to the Ti $t_{2g}$
character of the corresponding band. In spite of the small overlap
with energetically higher bands of predominant La $d$ character, the
Ti $t_{2g}$ bands are well-defined, and in particular there is a
rather sharp upper energy boundary, above which the Ti $t_{2g}$
character is negligible.

The comparison of the Ti $t_{2g}$ bands for compressive
(Fig.~\ref{fig:bands}a) and tensile (Fig.~\ref{fig:bands}c) strain
shows that there are some noticeable changes in the band dispersion,
but in all cases the Ti $t_{2g}$ bands remain well-defined within an
energy window between $-$1.0\,eV and $\sim$1.5\,eV around the Fermi
level. The overall Ti $t_{2g}$ bandwidth is slightly reduced
(enlarged) for tensile (compressive) strain compared to the unstrained
case.

\paragraph*{Construction of maximally localized Wannier functions.}

Similar to previous DMFT studies of LaTiO$_3$
\cite{Pavarini_et_al:2004, Craco_et_al:2004, Pavarini_et_al:2005} we
focus on the bands with predominant Ti $t_{2g}$ character around the
Fermi energy. In order to construct a set of maximally localized
Wannier functions \cite{Marzari_et_al:2012} for these bands, we define
an energy window for each strained structure based on the
corresponding Ti $t_{2g}$ ``fatbands''. The upper boundary of this
energy window is defined as the energy above which the Ti $t_{2g}$
character becomes negligible (see dashed blue lines in
Fig.~\ref{fig:bands}). We have verified that small changes (of the
order of $\sim$0.1-0.2\,eV) of this upper boundary have no significant
effects on our results.

\begin{figure}
\flushleft
\raisebox{1.9cm}{a)\hspace*{0.65cm}}
\begin{minipage}[c]{0.54\columnwidth}
\includegraphics[width=\textwidth]{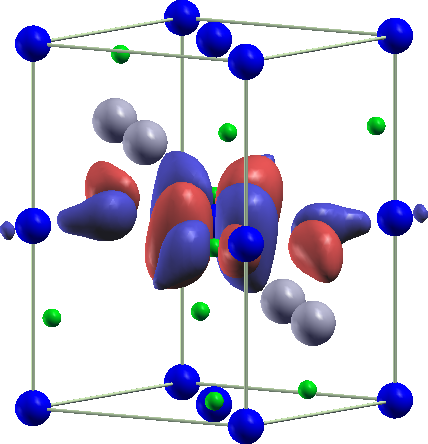}
\end{minipage}

\smallskip

\raisebox{1.9cm}{b)\hspace*{0.8cm}} 
\begin{minipage}[c]{0.5\columnwidth}
\includegraphics[width=\textwidth]{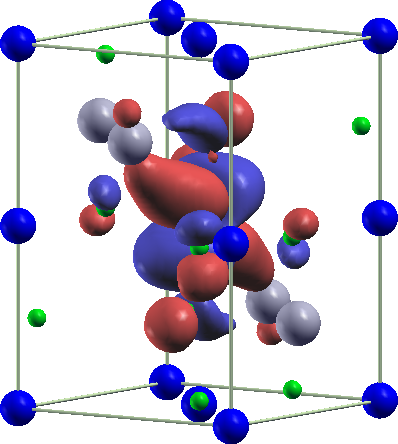}
\end{minipage}

\smallskip

\raisebox{1.9cm}{c)\hspace*{0.8cm}}
\begin{minipage}[c]{0.5\columnwidth}
\includegraphics[width=\textwidth]{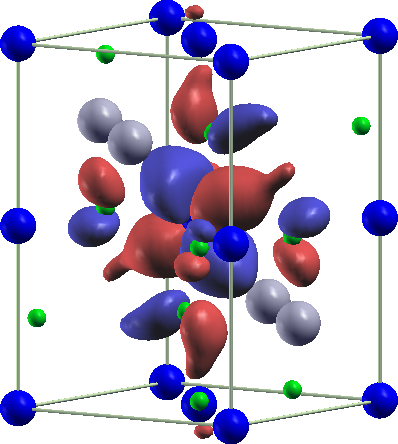}
\end{minipage}

\begin{picture}(0,0)(-185,-190)
\thicklines
\put(0,0){\vector(0,1){40}}
\put(0,0){\vector(1,0){35}}
\put(0,0){\vector(-4,3){15}}
\put(1,43){$z$}
\put(32,5){$x$}
\put(-16,16){$y$}
\end{picture}
\caption{The three maximally localized Wannier functions located on
  the same Ti site, calculated for the nominally unstrained
  structure. From top to bottom the corresponding Wannier functions
  are obtained by initial projection on $d_{xy}$ (a), $d_{yz}$ (b),
  and $d_{xz}$ (c) orbitals, respectively.}
\label{fig:wannier}
\end{figure}

We then use the wannier90 code \cite{Mostofi_et_al:2008} to obtain
three maximally localized Wannier functions per Ti site, starting from
initial projections on atomic Ti $t_{2g}$ orbitals. The resulting
Wannier functions for the nominally unstrained case are depicted in
Fig.~\ref{fig:wannier}. It can be seen that the resulting Wannier
functions resemble atomic $d$ states with $t_{2g}$ character centered
on the Ti sites, but also exhibit strong $p$-like tails situated at
the surrounding oxygen sites. Overall, the orbital character is less
obvious compared to similar $t_{2g}$ Wannier functions for SrVO$_3$
(see e.g. Ref.~\onlinecite{Lechermann_et_al:2006}). This is due to the
tilts and rotations of the oxygen octahedra surrounding the Ti
sites. However, there are strong similarities between our Wannier
functions and the $N$'th order muffin-tin orbitals calculated for
LaTiO$_3$ in Ref.~\onlinecite{Pavarini_et_al:2005}.

\paragraph*{Nearest neighbor hopping amplitudes.}

\begin{figure}
\raisebox{2.3cm}{a)}
\begin{minipage}[c]{0.85\columnwidth}
\includegraphics[width=\textwidth]{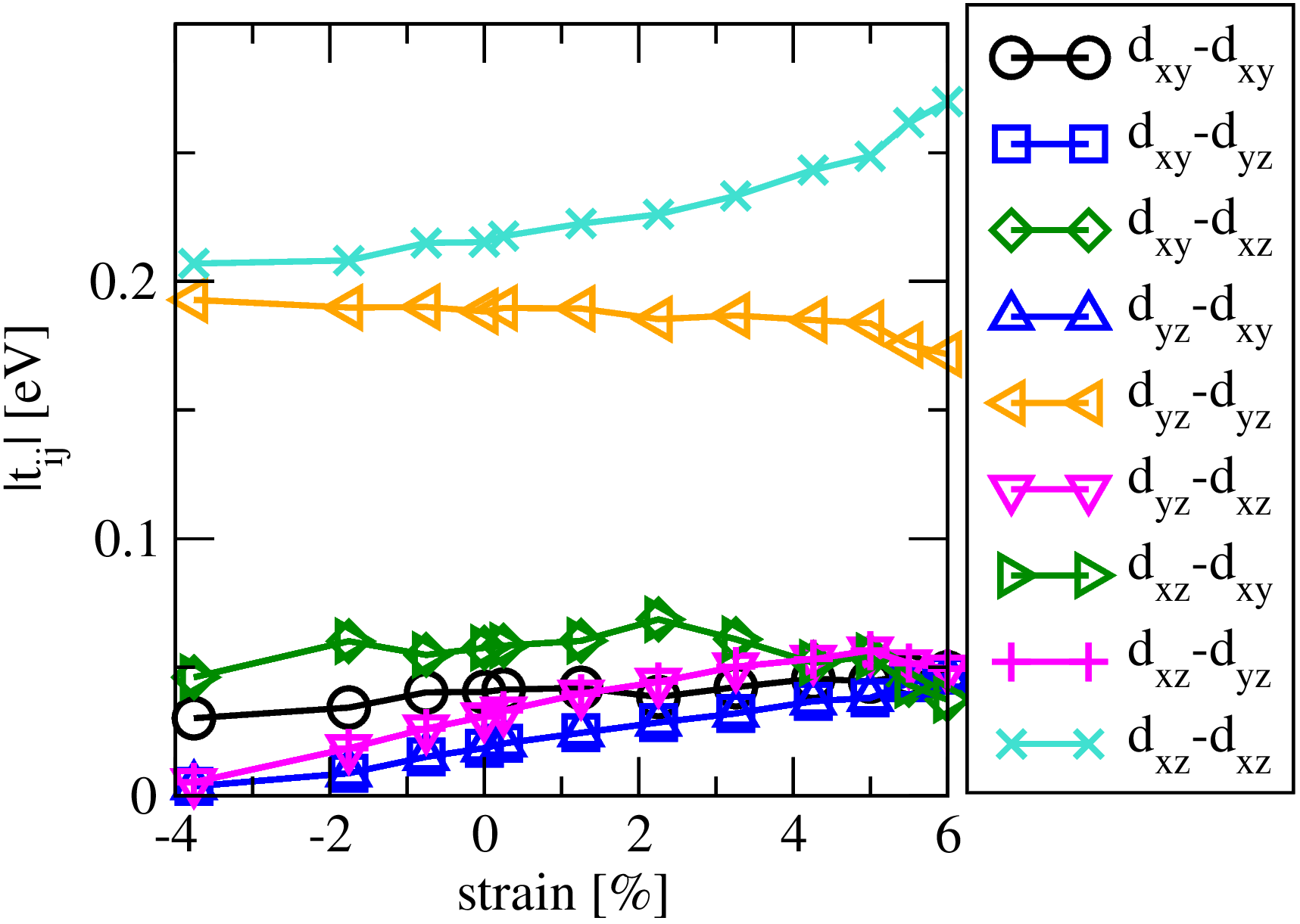}
\end{minipage}

\smallskip

\raisebox{2.3cm}{b)} 
\begin{minipage}[c]{0.85\columnwidth}
\includegraphics[width=\textwidth]{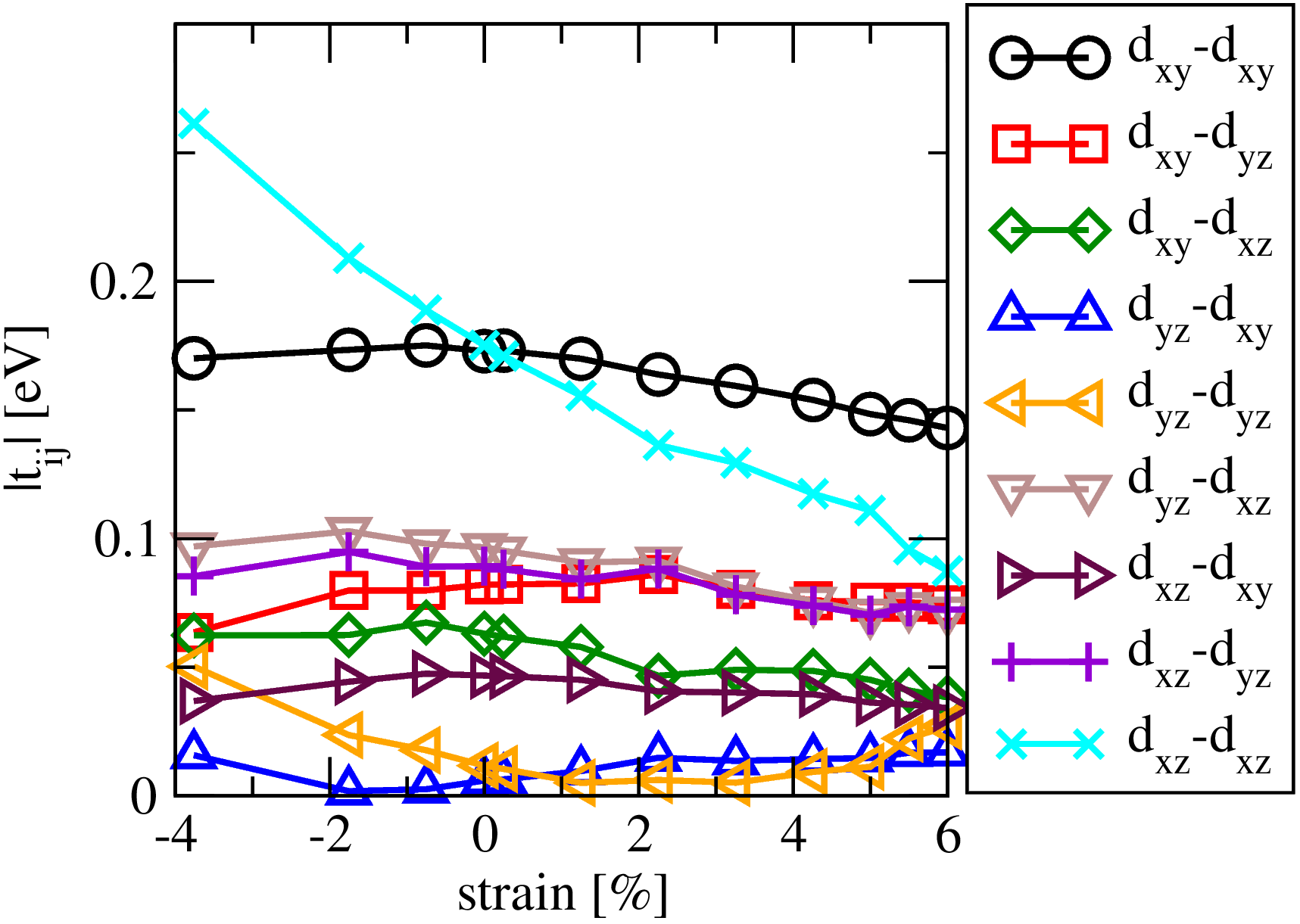}
\end{minipage}
\caption{Nearest neighbor hopping amplitudes, corresponding to hopping
  between effective Ti $t_{2g}$ Wannier functions, as function of
  epitaxial strain. a) hopping along the orthorhombic $c$
  direction. b) hopping within the $a$-$b$ plane. The orbital
  character that we use to denote the various hopping amplitudes
  refers to the initial projections used to obtain the corresponding
  Wannier functions. As can be seen from Fig.~\ref{fig:wannier}, the
  actual orbital character of the resulting Wannier function can
  exhibit significant deviations from this initial projection.}
\label{fig:hopping}
\end{figure}

Fig.~\ref{fig:hopping} shows the hopping amplitudes between
neighboring Ti sites as a function of strain. For the out-of-plane
hopping the $d_{xz}$-$d_{xz}$ and $d_{yz}$-$d_{yz}$ hopping amplitudes
are noticeably larger than all other hopping amplitudes. This, to some
extent, resembles the case of the perfect cubic perovskite structure,
where these two hoppings are the only ones that are allowed by
symmetry. However, no clear trend can be observed as function of
strain. The $d_{xz}$ hopping increases with increasing (tensile)
strain, consistent with the smaller lattice constant along $c$, but
simultaneously the $d_{yz}$ hopping decreases. For the in-plane
hoppings, a strong decrase is visible for the $d_{xz}$-$d_{xz}$
hoppings with increasing tensile strain, but all other hopping
amplitudes show only a weak strain dependence. These conflicting
trends can partially be explained by the complicated changes in the
orbital character as function of the octahedral tilts and strain, but
also reflect the fact that the trends expected from the changes in
bond distances are opposite to the trends expected from the bond
angles. For compressive strain, the shorter in-plane bond distances
would in principle enhance the corresponding hopping amplitudes. On
the other hand, the simultaneous increase of the angle $\theta$ leads
to a stronger distortion of the in-plane Ti-O-Ti bond angle, which is
detrimental to the hopping. Thus, the trends expected from the changes
in bond distances and bond angles are opposite to each other and it is
unclear how the hopping between the $t_{2g}$ states at neighboring Ti
sites will be affected by the structural changes described in the main
article.


\paragraph*{Further details of our DMFT approach.}

To account for effects of the electron-electron interaction within the
partially filled $t_{2g}$ bands, we use dynamical mean-field theory
(DMFT), where the local Green's function corresponding to the
effective Ti $t_{2g}$ states is calculated using a
momentum-independent self-energy:
\begin{equation}
  G_\text{loc}(\text{i}\omega_n)=\frac{1}{N_k}\sum_{\mathbf{k}}\left[
    \text{i}\omega_n+\mu-H_0(\mathbf{k})-\Sigma(\text{i}\omega_n)
    \right]^{-1}\,.
\label{eq:gloc}
\end{equation}
The self-energy $\Sigma(\text{i}\omega_n)$ is determined through a
mapping to a fictitious impurity problem with $G_\text{imp} =
G_\text{loc}$ and the same local interaction as the original lattice
problem.\cite{Georges_et_al:1996} The noninteracting Hamiltonian
$H_0(\mathbf{k})$ in Eq.~(\ref{eq:gloc}) corresponds to the Kohn-Sham
Hamiltonian expressed in terms of our maximally localized Wannier
functions with Ti $t_{2g}$ character. The corresponding full
Hamiltonian is then given by $H = H_0 + H_\text{int}$, where we use
the so-called Slater-Kanamori form for the local electron-electron
interaction, including both spin-flip and pair-hopping terms:
\begin{align*}
&H_\text{int}=\sum_{a} U n_{a,\uparrow} n_{a,\downarrow}+\sum_{a\ne
    b,\sigma} U' n_{a,\sigma} n_{b,-\sigma} \nonumber\\ &+\sum_{a\ne
    b,\sigma} (U'-J) n_{a,\sigma}n_{b,\sigma}\nonumber\\ &-\sum_{a\ne
    b}J(d^\dagger_{a,\downarrow}d^\dagger_{b,\uparrow}d_{b,\downarrow}d
  _{a,\uparrow} +
  d^\dagger_{b,\uparrow}d^\dagger_{b,\downarrow}d_{a,\uparrow}d_{a,\downarrow}
  + h.c.) \ ,
\end{align*}
where $d^\dagger_{a,\sigma}$ is the creation operator for an electron
in Wannier orbital $a$ with spin $\sigma$,
$n_{a,\sigma}=d^\dagger_{a,\sigma}d_{a,\sigma}$, and $U'=U-2J$.

The full interacting Green's function of the fictitious impurity
problem is then calculated using a continuous time hybridization
expansion quantum Monte-Carlo solver \cite{Gull_et_al:2011}
implemented in the TRIQS 0.9 code,\cite{triqs} which uses a
representation of the Green's function in terms of Legendre
polynomials.\cite{Boehnke_et_al:2011} Legendre coefficients with $l >
40$ are set to zero. Spectral functions are calculated from the
imaginary time Green's function using the maximum entropy method
(classical MaxEnt).\cite{Jarrell/Gubernatis:1996}

\bibliography{references}